\documentclass[fleqn,10pt]{wlscirep}
\usepackage{bbm,multicol,graphicx,float}

\newcommand{\Niso}{$^{14}$N }
\def\be{\begin{equation}}
\def\ee{\end{equation}}
\def\e#1{\label{#1}\end{equation}}
\def\bea{\begin{eqnarray}}
\def\eea{\end{eqnarray}}
\def\ea#1{\label{#1}\end{eqnarray}}

\def\bem#1{\begin{mathletters}\label{#1}}
\def\eml{\end{mathletters}}

\def\ket#1{{|#1\rangle}}

\def\ketbra#1{{|#1\rangle}{\langle#1|}}

\def\4#1{{\boldsymbol{#1}}}
\def\8#1{{\widetilde{#1}}}
\def\bse{\begin{subequations}}
\def\ese{\end{subequations}}
\def\nn{\nonumber}
\title{Purification of an unpolarized spin ensemble into entangled singlet pairs}

\author[1]{Johannes N. Greiner}
\author[1,2,*]{Durga Bhaktavatsala Rao Dasari}
\author[1,2]{Jörg Wrachtrup}
\affil[1]{3rd Institute of Physics, University of Stuttgart, Pfaffenwaldring 57, 70569 Stuttgart, Germany.}
\affil[2]{Max Planck Institute for Solid State Research, Heisenbergstraße 1, 70569 Stuttgart, Germany}
\affil[*]{d.dasari@physik.uni-stuttgart.de}

\begin{abstract}
Dynamical polarization of nuclear spin ensembles is of central importance for magnetic resonance studies, precision sensing and for applications in quantum information theory. Here we propose a scheme to generate long-lived singlet pairs in an unpolarized nuclear spin ensemble which is dipolar coupled to the electron spins of a Nitrogen Vacancy center in diamond. The quantum mechanical back-action induced by frequent spin-selective readout of the NV centers allows the nuclear spins to pair up into maximally entangled singlet pairs. Counterintuitively, the robustness of the pair formation to dephasing noise improves with increasing size of the spin ensemble. We also show how the paired nuclear spin state allows for enhanced sensing capabilities of NV centers in diamond.
\end{abstract}
\begin{document}

\flushbottom
\maketitle
\thispagestyle{empty}

\section*{Introduction}
The induction of quantum correlations between any two physical systems interacting with a common environment is a well-studied mechanism in the research of strongly correlated systems. Singlet pairing among electrons mediated by phonons in superconductors\cite{anderson_resonating_1987,frohlich_theory_1950} or pairing among next nearest neighbors induced by superexchange interaction are well known examples of such mechanisms\cite{anderson_antiferromagnetism._1950}. Quantum particles paired in singlets share maximal entanglement and their total vanishing spin additionally renders them robust to external noise\cite{carravetta_beyond_2004}. Singlets are also known to form ground states of certain spin-$1/2$ dimer compounds \cite{singh_singlet_2007}. As singlets have zero spin and behave like bosons, systems with a spin singlet ground state have also been extensively studied in the context of Bose-Einstein condensation\cite{nikuni_bose-einstein_2000}. Thus pairing spins into singlet states is relevant for understanding both many body aspects of spin dynamics in strongly correlated systems and for quantum protocols that require long spin life times ($T_1$) \cite{levitt_singlet_2012}. For example such extended life times of over few hours are observed for protons in parahydrogen, when paired as singlets, making them a natural resource for polarization experiments. Singlets are also a key resource for entanglement and quantum information protocols\cite{bennett_purification_1996,gisin_spin_1999,chakraborty_probing_2013,behbood_generation_2014}. In this work we show how unpolarized nuclear spins dynamically pair (form spin-$1/2$ dimers) into maximally entangled (singlet) states when they are inhomgoeneously coupled to an electron spin that is repeatedly projected onto a preferred state.

Optically addressable electron spins associated with atomic-sized nitrogen-vacancy (NV) defect centres in diamond have been extensively studied over the past decade for precision sensing, and for implementing quantum information protocols at ambient conditions. With the ability to achieve near unity polarization on very short time scales, NVs can be used to polarize nearby nuclear spins achieving dynamical nuclear polarization even for low magnetic fields and high temperatures.\cite{alvarez_local_2015,chen_optical_2015} While the polarization transfer between electron and nuclear spins is well known in magnetic resonance studies, we show here that by further exploiting the state specific readout and polarization of the electron spins of NV centers it is possible to go beyond standard polarization schemes and achieve deterministic singlet pairing among randomly distributed nuclear spins. 

The NV center has an electronic spin triplet ($S=1$) ground state and an intrinsic \Niso  nuclear spin. The center is surrounded by a nuclear spin bath of ${}^{13}C$ spins  with a natural abundance of about $1.1$ percent. The NV can be optically excited by a strong off-resonant laser allowing it to polarize into one of the ground states. On the other hand at low temperature $(4K$) where the excited state spectrum of the NV is well resolved, it can be excited by resonant optical fields. Due to this, the three ground states can be individually read out and initialized with high fidelity.\cite{dreau_single-shot_2013} Such state specific excitation and readout has already been used to achieve deterministic nuclear spin state preparation \cite{togan_laser_2011}, entanglement between solid state qubits mediated by photons \cite{bernien_heralded_2013}, transfer and storage of single photon states in nuclear spins \cite{yang_high-fidelity_2016}, cooling nanomechanical oscillators \cite{rao_heralded_2016} and for generating decoherence free subspaces \cite{kalb_experimental_2016}. Entanglement induced by projective measurement has also been considered in electron scattering\cite{costa_entanglement_2006} and for models in condensed matter and quantum optical systems\cite{wu_long-range_2004}. While previous studies are either restricted to fewer quantum subsystems or to a smaller number of projections, we here give a formal prescription for the dynamics induced on an arbitrarily sized spin ensemble by multiple projective measurements of a central spin.
\begin{figure}[H]
\centering
\includegraphics[width=.76 \textwidth]{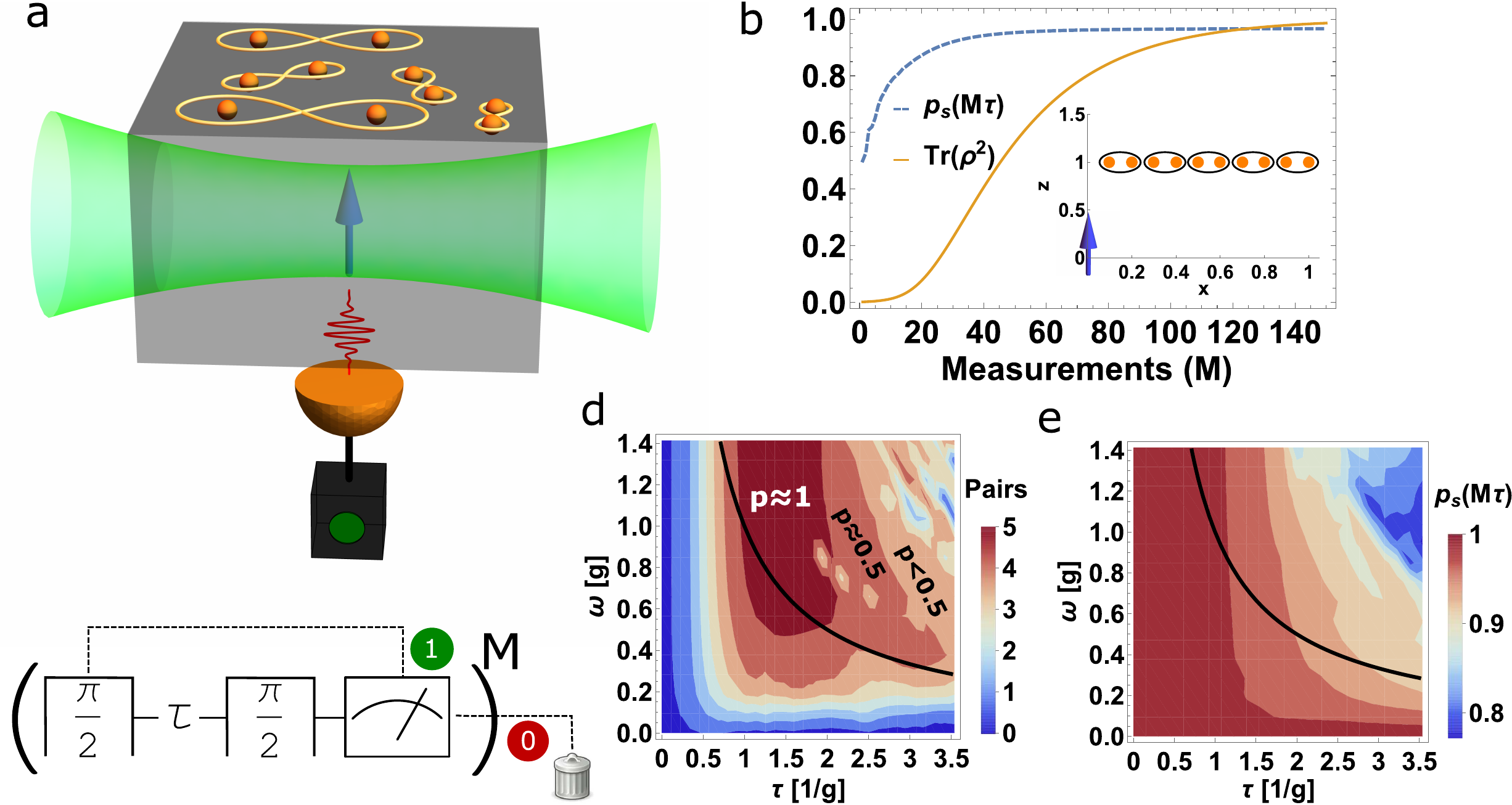}
\caption{(a) Schematic illustration of the scheme, where the central spin (single NV center in diamond) shown as solid blue arrow, is dipolar coupled to nuclear spins (orange spheres) on the surface. The spin is initialized and read out optically.  Successful state selective readouts lead to pairwise entanglement of the nuclear spins. (b) The purity and the success probability as a function of the number of measurements $M$ is shown for a spin ensemble of $10$ spins. The inset shows the linear chain of nuclear spins considered for this calculation. (c) Schematic of the pulse sequence and measurement on the NV electron spin. The $\omega-\tau$ contour plots in units of the effective coupling $ g=\sqrt{1/N \sum_k |\vec{g}_k|^2} $ are illustrating (d) the total number of singlet pairs and (e) the final success probability $P_S(M \tau)$. In (d) we also show the corresponding purity $ p $ of the spin bath. Moreover the line $ \omega=1/\tau $ is plotted in (d) and (e) to indicate a good choice of time and frequency for both pairing and success probability.}
\label{fig:probabilitypurity}
\end{figure}
\begin{multicols}{2}
\section*{Results}

\subsection*{Model}
We consider a generic central spin model with a central spin-$S$ (spin-$1/2$) inhomgoeneously coupled to a spin-bath composed of spin-$1/2$ particles ($I$). One example of such a central spin system is given by choosing any two-level subspace within the triplet ground state of an NV electron spin in diamond. The spin-bath is then constituted either by the internal ${}^{13}C$ or by external nuclear spins. Due to a very weak dipolar coupling among the nuclear spins, we consider the spin-bath to be noninteracting. In the reference frame of the central spin (NV) one can think of the spins of the bath as being randomly distributed in a three-dimensional plane comprising of the $S$-spin polarization and the plane perpendicular to it. 
With the magnetic field aligned along the central spin axis (say NV $z$-axis) the Hamiltonian that determines the dynamics is given by\cite{doherty_nitrogen-vacancy_2013} 
\begin{equation}
H = S^z \otimes \sum_k \vec{g}_k (r) \cdot \vec{I}_k + \omega \sum_k \mathbbm{1} \otimes I_z^{(k)}
\label{eq:totalHamiltonian}
\end{equation}
where $g_k$ is the strength of dipolar coupling between the central spin and the $k$-th nuclear spin, dependent on the spatial separation between these spins. The random spatial location of the spins with respect to the probe results in inhomogeneous couplings such that the bath has neither conserved quantities nor preferred symmetries. The external field $\omega$ under which the nuclear spins precess is assumed to be uniform over the entire sample and is along the $z$-direction. The dynamics generated by the above Hamiltonian is exactly solvable \cite{bhaktavatsala_rao_controlled_2007}.
In the $S^z$ basis of the central spin ($|{1}\rangle, |-1\rangle$), the above Hamiltonian can be rewritten as 
\bea
H &=& H^+|1\rangle\langle 1| + H^-|-1\rangle\langle -1|, \nn \\
H^{\pm} &=& \sum_k \omega_k I^k_z \pm \sum_k \vec{g}_k (r) \cdot \vec{I}_k
\label{eq:HPlusMinus} 
\eea

where $H^{\pm}$ are the nuclear spin-bath operators. 
The simple form of $H^{\pm}$ makes it easy to diagonalize, so that we obtain a closed-form equation for the time-evolution operator of the total system
\be
\label{uop}
U(t) =\left[U^+_I(t)|1\rangle\langle 1| \, + \, U^-_I(t)|{ -1}\rangle\langle-1|\right]
\ee
where $U^\pm_I(t) = \bigotimes_k U^\pm_k(t)$, and 
\be
\label{udef}
 U^\pm_k(t) = \left [\cos \delta_k t \hat{\rm I}+ \frac{i\sin \delta_k t}{\delta_k}(\omega I^z_k \pm\vec{g}_k (r) \cdot \vec{I}_k) \right].
\ee
with $\delta_k = \sqrt{\omega^2+|\vec{g}_k(r)|^2}$.

\subsection*{Analysis}
 Starting from the initial spin state $\ket{-1}$, we prepare a superposition state using the first $\pi/2$ pulse and allow it to evolve for a time $\tau$. The spin is then brought back to the energy basis ($\ket{1(-1)}$) with the final $\pi/2$ pulse. We now read out the state of the spin optically and note the measurement result. Conditional on a successful measurement result, e.g. finding the spin in the state $\ket{-1}$, the procedure is repeated $M$ times, as shown in Fig.1. 
Since the time $\tau$ is arbitrary, the probability of obtaining a successful measurement is always below unity. Given this conditional spin state dynamics, the nuclear spin ensemble evolves from a fully mixed state to a pure state as we show below.

A successful spin state measurement projects the spin-bath onto a state
\be
\rho_I(\tau) = \frac{V(\tau)\rho_I(0)V(\tau)^\dagger}{{\rm Tr}[V(\tau)\rho_I(0)V(\tau)^\dagger]}.
\ee
where $V(\tau) = \frac{1}{2}\left(U^+_I + U^-_I\right)$, and $\rho_I(0)$ is the initial state of the spin-bath. As $V(\tau)$ is a non-unitary operator, the state needs to be normalized as shown above to get a valid density matrix for the spin-bath. The success probability for this projection is then simply given by 
\bea 
P_S(\tau) = {{\rm Tr}[V(\tau)\rho_I(0)V(\tau)^\dagger]}.
\eea 
Now $\rho_I(\tau)$ is the initial state of the spin-bath for the next repetition and after evolving again for a time $\tau$, a successful measurement of the spin will project the spin-bath onto the state $\rho_I(2\tau) = \frac{V(\tau)\rho_I(\tau)V(\tau)^\dagger}{{\rm Tr}[V(\tau)\rho_I(\tau)V(\tau)^\dagger]}$. 
After $M$ successful projections of the spin to its desired state, the final state of the spin-bath is 
given by
\be
\rho_I(t=M\tau) = \frac{V^M(\tau)\rho_I(0)V^M(\tau)^\dagger}{{\rm Tr}[V^M(\tau)\rho_I(0)V^M(\tau)^\dagger]}.
\ee
If $\rho_I(t)$ is the desired state of the dynamics, the probability to achieve this state is given by
\bea 
P_S(t) = \prod_{k=1}^MP_S(k\tau),
\eea
which is usually much smaller than unity. To obtain the final state given in Eq.(7) within the relaxation time of the spin-bath, the probability of obtaining successful projections should slowly converge to unity such that $P_S(t)$ attains a constant value and does not change with further measurements.\cite{nakazato_purification_2003} Hence, the readout times $\tau$ and the external field $\omega$ have to be chosen optimally such that both the overlap with the desired end state and the total probability $P_S(t)$ is maximized. After the above discussion on the conditions required to observe measurement induced back-action on the dynamics of the spin bath, we will now characterize the steady state. 

\subsection*{Dynamics}
The conditional dynamics imposed by the central spin $S$ on an unpolarized bath allow the generation of both classical and quantum correlations among the bath spins alongside the purification of the spin ensemble. In the steady state the nuclear spins can either be (i) purely uncorrelated, $\rho_U = \rho_1\otimes\rho_2\cdots\otimes\rho_n$, (ii) classically correlated, $\rho_C = \sum_k p_k \rho_{1,k}\otimes\rho_{2,k}\cdots\otimes\rho_{n,k}$ or (iii) quantum mechanically correlated in which case the total nuclear spin state $\rho_Q$ cannot be written as either of the above. Using the measurement based protocol described above one can tune the external field to generate either of these cases. 
\subsubsection*{Classical correlations}
 To observe only classical correlations one may switch off the external magnetic field, i.e. set $\omega=0$. In this case, $U^+_I U^-_I  =\hat{I}$, and the non-unitary operator $V$ takes a simple form, $V(\tau) = {\rm Re}[U^+_I]$. As we start with a completely unpolarized bath, we can choose the bath spins to be diagonal in the eigenbasis of  $H_B = \sum_k \vec{g}_k (r) \cdot \vec{I}_k$. Since there is no magnetic field, the spin bath, which is diagonal in the eigenbasis of $H_B$
initially, continues to remain diagonal throughout the evolution. As $U^{\pm}_I$ is diagonal in the eigenbasis of $H_B(t)$, so is the operator $V$. 
In the limit of large number of measurements, the operator $V^M$ has only one surviving diagonal entry which corresponds to the maximum eigenvalue of $V$.\cite{nakazato_purification_2003}  The gap between the maximum eigenvalue with the remainder sets the speed of convergence towards the steady state. The maximum value is chosen both by the free evolution time $\tau$ and the couplings $g_k$. For the case of identical couplings there are ${}^NC_{N/2}$ degenerate eigenvalues of $V$ that attain the maximum value independent of $\tau$. Due to this, the bath gets polarized only into the sector for which $\langle \sum_k I^z_k \rangle = 0$. Hence, the enhancement in the purity is only by a factor $\frac{2^N}{{}^NC_{N/2}}$.  On the other hand for completely random couplings the maximum eigenvalue sector is doubly degenerate, thereby allowing the bath to achieve a purity of $0.5$. In the limit of large number of measurements $M$, the steady state of the bath
\bea
\rho_C = \frac{1}{2}\left(\ketbra{\psi_{+}} + \ketbra{\psi_{-}} \right)
\eea
is an equal mixture of the macroscopic superposition states $\ket{\psi_{\pm}} = \frac{1}{\sqrt{2}}\left[\ket{\phi_1\phi_2\phi_3\cdots\phi_n}\pm\ket{\phi^\perp_1\phi^\perp_2\phi^\perp_3\cdots\phi^\perp_n}\right]$, where $\ket{\phi_i},\ket{\phi^\perp_i}$ is the diagonal basis of any $i$-th spin. Clearly, even though the state is mixed it cannot be written as a product state of individual spins, and hence is classically correlated.
\subsubsection*{Quantum correlations - Singlet Pairing}
By switching on the magnetic field ($\omega \ne 0$), the unitary operators $U^+_I$, $U^-_I$ can no more be diagonalized in a common eigenbasis as the two parts of the Hamiltonian given in Eq. (1) do not commute. Due to the mixing caused by the applied field in the eigenbasis of $H_B$, powers of the operator $V$ cannot be diagonalized in a given basis. 
\begin{figure*}
\centering
\includegraphics[width=.76\textwidth]{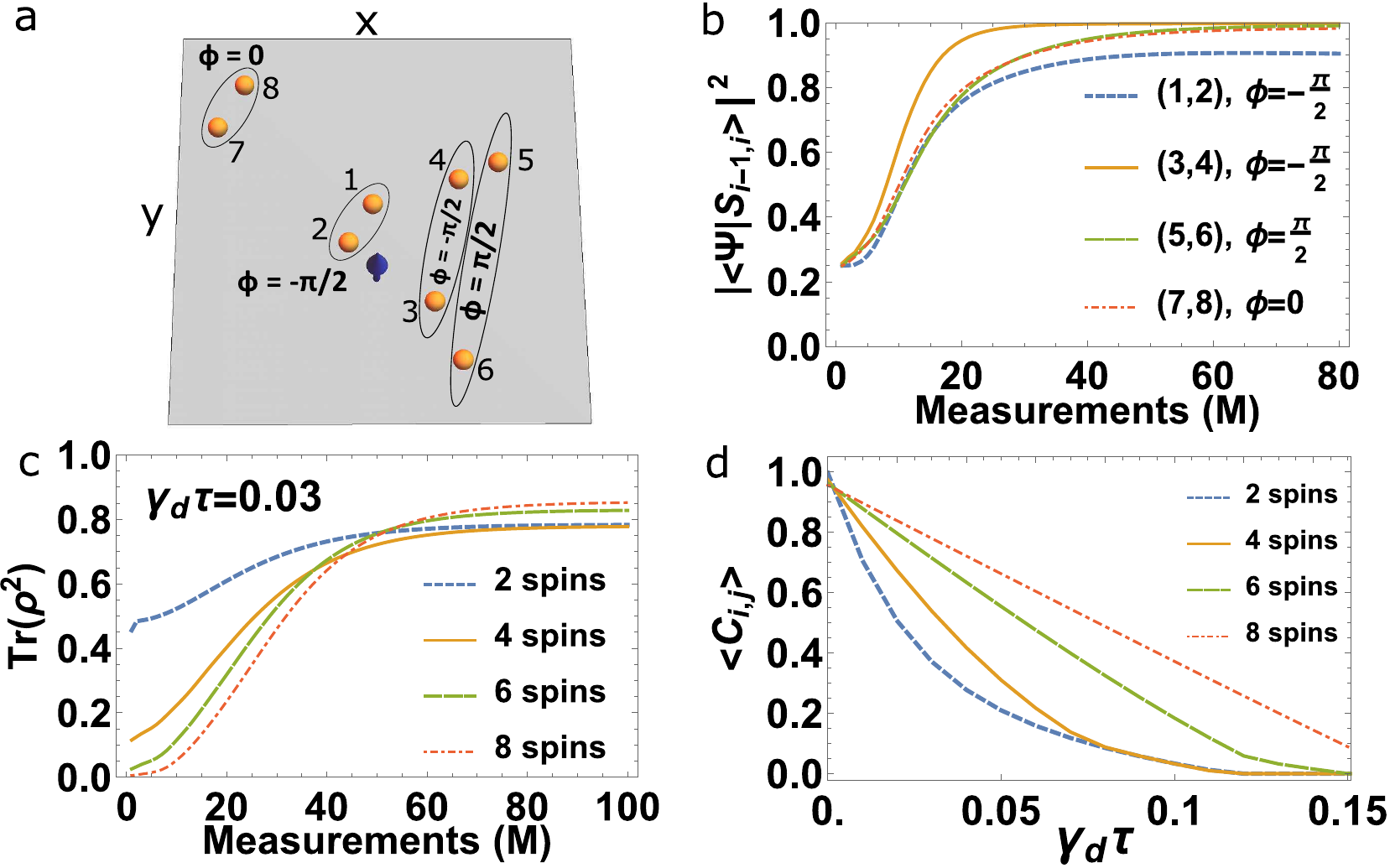}
\caption{(a) Nuclear spins are randomly distributed in the $ x-y $ plane and pair up into the entangled state $ \ket{\tilde{S}_{i,j}} = \frac{1}{\sqrt{2}}[\ket{1,-1}-{\rm e}^{i\phi}\ket{-1,1}]$. We also show the relative phase $\phi$ of each pair in this configuration.  (b) We plot the entanglement fidelity for various pairs as function of measurements $M$. (c) The purity is shown as a function of the number of measurements for different spin numbers in the presence of dephasing. (d) The average pairwise concurrence of the steady state is plotted as a function of the dephasing rate $\gamma_d$ multiplied by $ \tau  $. }
\end{figure*}
This non-trivial structure of $V^M$ results in generating quantum correlations 
among the baths spins with increasing $M$.\cite{nakazato_purification_2003} For random couplings $g_k$, the dynamics is no more amenable to analytical diagonalization. We hence resort to exact numerical diagonalization methods and develop a technique to simulate spin baths with larger number of spins. 
We shall first consider a linear chain of spins arranged along the $x$-direction at a distance $z_0$ along the $z$-axis, from the central spin $S$. Though the geometry 
is one-dimensional and the spins are positioned homogeneously, the $ 1/r^3 $ dependence of the dipolar coupling results in nonlinear variation in both $x$ and $z$ components of the coupling $g$. Due to this inhomogeneity, for any $k$-th spin the spin to the left ($k-1$) appears more homogeneous in coupling than the spin to the right ($k+1$). This marks the key step for singlet pairing. In Fig. 1b we plot the purity and measurement probability as a function of increasing number of measurements. The bath evolves towards a pure state with a high probability, and the spins of the bath pair up in singlets as shown in Fig. 1a. 
As mentioned above the singlet pairing happens only between the $k$-th and $(k-1)$-th spin which are less inhomogeneous in their coupling to the central spin $S$. Hence in the asymptotic limit, the steady state of the ensemble takes the form 
\bea
\ket{\psi}_Q = \bigotimes_{k=2}^N\ket{S_{k-1,k}},
\eea
where $\ket{S_{k-1,k}} = \frac{1}{\sqrt{2}}[\ket{1,-1}-\ket{-1,1}]$, written in the eigenbasis $\ket{\pm1}$, of the $I^z_k$ operator. While the partners of a given pair, here the $k$-th and $(k-1)$-th spins, are all unidentical with respect to the central spin, they still tend to pair up identically into a similar singlet state. This is remarkable, as inhomogeneity drives the ensemble towards a homogeneous state of identical singlet paired states. We would also like to note that such singlet pairing does not occur if the central spin were coupled identically to all other spins. We show in Fig. 1c,d the optimal conditions on the applied field and the free evolution time $\tau$ for obtaining singlet pairing with high probability. From the $\omega-\tau$ plot we see that not all values of applied fields $\omega$ and measurement times $\tau$ result in singlet pairing. We numerically find the optimal choice for the external field $\omega$ and the time $\tau$ required to purify the ensemble should be chosen such that
\bea
\tau \sim \frac{1}{\omega}, ~\omega \sim \frac{1}{2N} \sum_{l\in\lbrace x,y,z\rbrace} \sum_{k=1}^N|g^l_k|.
\eea

We now move to a more generic case where the spins are randomly distributed over the $x-y$ plane as shown in Fig. 2a. For complete random positioning of the nuclear spins there is still high probability to observe pairing of spins into entangled states. But now the entangled pair state is not only restricted to the singlet sector as it is the case for the $1D$ chain, but rather into the entangled state $ \ket{\tilde{S}_{i,j}} = \frac{1}{\sqrt{2}}[\ket{1,-1}-{\rm e}^{i\phi}\ket{-1,1}] $. 
The additional phase $\phi$ arises when the difference in the $x$ and $y$ couplings of the pairing spins to the central spin $S$ has a large variation. But we show the pairing in the real space and in the coupling space of a spin ensemble comprised of $8$ spins. 
The choice of pairing is purely determined by the minimal inhomogeneity of the partners in their coupling to the central spin. For example in the 1D-case (see Fig. 1b), the inhomogeneity decreases for farther away spins from the central spin. Hence pairing starts from the end of the chain farther away from the central spin and defines the pairing pattern in the chain. In the 2D-case the situation is more complex as the inhomogeneity both in the $x$ and $y$ direction together determine the pairing partners. Additionally, we see that singlets are not the only choice and there are four different relative phases $\phi = \lbrace{0, \frac{\pi}{2}, {\pi}, \frac{3\pi}{2}\rbrace}$ dependent on the $x-y$ position of the spins. For example in Fig. 2a, where spin $5$ could only pair with its farthest partner $6$, as all other spins have paired up. Remarkably, even though these spins are (effectively) weakly interacting, they still tend to polarize into entangled pairs rather than polarizing into some unentangled pure states.  
\begin{figure}[H]
\centering
\includegraphics[width=.4\textwidth]{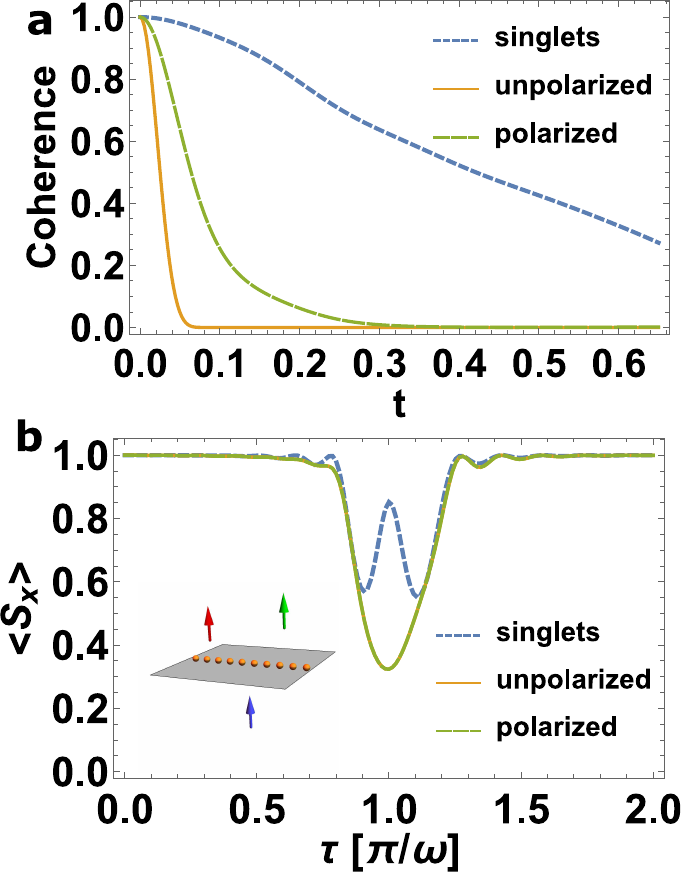}
\caption{(a) The coherence of the central spin is plotted as a function of time for three different initial states of the spin-bath. (b) The central spin transition conditioned on its coupling to three different nuclear spin species that have Larmor frequencies $\omega, \omega_1 = \omega + \epsilon, \omega_1 = \omega - \epsilon$ (we chose $ \epsilon = \omega/10 $) is shown as a function of the interrogating time $\tau$, for three different initial states of the strongly interacting spin-bath ($\omega$). Clearly, when the spin bath  is initialized into the singlet paired state, the two farther nuclear spins are well resolved. The inset displays the spin configuration with different spin species. }
\end{figure}
\subsection*{Dephasing of the nuclear spins}
During the cooling process, the nuclear spins can be subject to dephasing noise caused by the finite relaxation time ($T_1$) of the electron spin and due to the measurement of the electron spin. As $\tau \ll T_1$, the dephasing caused during the free evolution time is negligible, but on the other hand the dephasing occurring during the state selective readout of the central spin $S$ could have a dominant effect on the dynamics. For example, the electronic spins of NV centers in diamond are measured by polarization dependent resonant optical excitation. The difference in the ground and excited state dipolar coupling strengths between the electron spins will give rise to a random phase for the nuclear spins, causing them to dephase. We have also evaluated these dephasing effects (see Methods) on the pairing process for different sizes of the spin ensemble. We see that the purity and the average pairwise concurrence becomes robust to dephasing with increasing size of the ensemble. 
This is counterintuitive, as the scaling of the decoherence is known to become worse with increasing size of the system\cite{aolita_scaling_2008}. To understand this inverse behavior in the measurement-driven dynamics we analyze the process involved during the projection by the operator $V$. 
The act of projection is to change (increase) the purity of the spin-bath. The relative change in purity from $m$-th and $(m+1)$-th measurement depends on the size of the ensemble, and similarly dephasing also changes (decreases) the purity between  $m$-th and $(m+1)$-th measurement depending on the ensemble size. The ratio between the two competing processes decides the equilibrium value of the purity (see Fig. 2c) and the equilibrium value of entanglement (Fig. 2d) in the spin-bath. 

\subsection*{Enhanced Quantum Sensing}
The measurement-induced quantum state engineering of the spin-bath can be advantageous for quantum information protocols for sensing applications\cite{zaiser_enhancing_2016,greiner_indirect_2015,unden_quantum_2016,cappellaro_quantum_2009,urizar-lanz_macroscopic_2013}. The convergence of measurement probability towards unity is a signature of a decoherence free environment for the central spin. By deterministically pairing the right partners into maximally entangled states, a decoherence free environment is obtained for the central spin (see Fig. 3a). In addition to long coherence times, the central spin has become transparent to farther away spins which were initially unresolvable when the spin-bath was in an unpolarized state. We show this superresolution in Fig. 3b, where two spin species with different Larmor frequencies are better resolved when the closest (strongly interacting) spin-bath is prepared into a paired state. 

\section*{Discussion}
The proposed scheme can be experimentally implemented at low temperatures (T$<$8 K) in a low strain ($\approx$1.2 GHz) NV center.  At such temperatures the optical transitions are well resolved allowing for resonant excitation to perform efficient initialization and projective high-fidelity single shot readout on the electron spin (see Methods). For a nuclear spin ensemble (protons, fluorine \cite{cai_large-scale_2013}) deposited on the surface of the diamond with a NV center at a depth of $5$ nm, and having a $2D$ distribution (i.e. the variation of the spin density along the $z$ (principal axis of NV) is much smaller than that along $x$ and $y$), with a density of $\sim 5{\rm nm}^{-2}$, the average field required to observe pairing with high probability will be $\omega \sim 10^{-2}$T and the probing time will be $\tau \sim 2.5\mu$s. With a typical readout time of the electron spin $\sim1\mu$s, the time required for $100$ measurements would be $\sim 0.35$ms, which is well within the relaxation times of the nuclear spins. The total probability to achieve success in all $100$ measurements for the case of $8$ nuclear spins is estimated to be $\sim 10^{-3}$. Thus, the event rate to observe high fidelity ($0.95$) pair formation would be one event per $\sim 3$sec. Over such time scales the singlet pairs can decohere both due to the $T_1$ of the NV center and also due to the surrounding electron spin-bath formed of paramagnetic defects (P1) and surface spins. With a typical density of $10^{15} cm^{-3} \, $ \cite{jarmola_temperature-_2012,bar-gill_solid-state_2013} these spins should contribute to dephasing on the order of seconds, and can be removed by optimally choosing the NV centers and with better surface treatments of the diamond. Thus the dominant contribution would come from the $T_1$ of the NV center which can be of the order of $1 - 10$ seconds at liquid-He temperatures \cite{bar-gill_solid-state_2013}.

One may also generalize the proposed scheme to other physical systems based on central spin models, where the central spin can be projectively readout. For example using impurity mediated coupling of Rydberg-excited atoms\cite{saffman_quantum_2010} or superconducting qubits \cite{chen_qubit_2014} where projective measurements can be performed with very high fidelities\cite{riste_feedback_2012}.

To conclude, we have shown how nuclear spins can pair up into maximally entangled states as they are cooled down via a central spin, subject to frequent projective spin state measurements. The natural inhomogeneity in their coupling to the central spin imposed by the dipolar coupling is key for observing such pairing phenomena. While standard methods for polarizing nuclear spins by the NV center only use its fast optical pumping feature, we show that using its high fidelity state-selective readout allows to steer the quantum correlations among the nuclear spins alongside their polarization. Additionally, we have shown how classical correlations can be converted into quantum correlations or vice-versa by changing the applied field or the measurement time. To observe the effects on a realistic timescale, i.e. within the $T_1$ times of the nuclear spins, the selective spin state measurements of the NV are required to occur with high probability. For this we have identified a regime for the applied field and measurement time that maximizes both the probability and the fidelity of obtaining the macroscopic pair state. The scheme can also be extended further to larger spin ensembles (see Methods) and to observe the formation of triplets (maximally entangled states of 3 spins) as well as higher clusters by manipulating the external frequency $ \omega $ and the measurement time $\tau$. The growth of entanglement through the spin diffusion process caused by their mutual dipolar couplings could shed light on the dynamics of many body correlations in nuclear spin ensembles.  In addition the singlet paired states could be a test bed for quantum simulations of low energy excitations in spin models with exact dimer ground states \cite{kumar_quantum_2002}.

\section*{Methods}
\subsection*{Analytical derivation of the effective dynamics}
Using spectral decomposition of the total Hamiltonian in equation \eqref{eq:totalHamiltonian} and time evolution according to the Schr\"odinger equation as $ U = e^{i H t} $ we can write
\begin{equation}
U=\left|1\right>\left<1\right|e^{i H_{+} t}+\left|-1\right>\left< -1 \right| e^{i H_{-} t}
\end{equation}
Thus, we decompose the time evolution operator of the combined NV and nuclear spin Hamiltonian into operators acting on the nuclear spins, depending on the state of the electron spin. $ H_\pm $ are defined as in equation \eqref{eq:HPlusMinus}.
As an example, if the electron spin is initialized into the spin state
$\left| \Psi_{el} \right> = \frac{\left| 1 \right>+\left| -1 \right>}{\sqrt{2}}$ 
it is clear that the time evolution yields
\begin{equation}
U \left| \Psi_{el} \right> \left| \Psi_{ns} \right>
= \frac{1}{\sqrt{2}} \left( \left| 1 \right> e^{i H_+ t} 
+ \left| -1 \right> e^{i H_- t} \right) \left| \Psi_{ns}\right> .
\end{equation}
When a subsequent state specific readout of the electron spin is performed, this results in an effective operation on the nuclear spins
\begin{equation}
\left| \Psi_{el} \right> \left< \Psi_{el} \right| U \left| \Psi_{el} \right> \left| \Psi_{ns} \right>
= \left| \Psi_{el}\right>
\frac{e^{i H_+ t} 
+ e^{i H_- t}}{2} \left| \Psi_{ns}\right>.
\end{equation}
The effect of a successful projection into a superposition state is thus a non-unitary operation $ V $ with the property
\begin{equation}
V \neq \bigotimes_i U_{ns,i} \,,
\end{equation}
which demonstrates that V can be an entangling operation. Moreover, in line with the findings of Nakazato et al. \cite{nakazato_purification_2003}, the final state of the evolved system is exclusively given by the (left) eigenstate $ \left| u_{max} \right> $ corresponding to the unique maximal eigenvalue of $ V $, in the limit of a large number $ N $ of projective measurements of the NV electron. Therefore, the choice of the state $ \left| \Psi_{el} \right> $ the NV is projected into, defines the target state the surrounding nuclear spins will reach, regardless of their initial state. This presents a useful tool for engineering the multi-qubit state of the nuclear spins.  Generalizing the example above, projecting into a general state
$ \left| \Psi_{el} \right> = 
\frac{ \alpha \left| 1 \right>+ \beta \left| -1 \right>}{\sqrt{|\alpha|^2 + |\beta|^2}} $
yields the operation on the nuclear spins as
\begin{equation}
V = |\alpha|^2 e^{i H_+ t} + |\beta|^2 e^{i H_- t} .
\end{equation} 
where $ |\alpha|^2 + |\beta|^2 = 1 $. By optimizing these degrees of freedom, a large array of target states defined by the respective operator $ V $ can be accessed. 
This includes states containing bipartite entanglement as presented in this manuscript, but generalizations to multipartite entanglement are possible. 
\subsection*{State selective readout of NV centers in diamond}
The electron spins of the NV center have a triplet ground state with magnetic quantum numbers $m_s = 0, ~\pm 1$. Spin selective initialization and readout of the ground state spins is optically possible only by resonant excitation at sufficiently low temperature (T$<$8 K), where the excitation spectrum of the NV center is well resolved \cite{togan_laser_2011,robledo_high-fidelity_2011,maze_properties_2011}. One can induce a cycling transition between the ground   $\left| m_s = 0\right> $ and the excited states $ \left| E_x \right> $ or $ \left| E_y \right> $ to achieve a high fidelity singlet shot (projective) readout of the NV center in the state  $\left| m_s = 0\right> $. Due to the high probability of spin mixing in the excited state, this readout leads to large dephasing of nuclear spin correlations. To avoid this, one can make use of another resonant excitation channel i.e., from the ground $\left| m_s = \pm 1\right> $ to the excited states $  \left| A_{1(2)} \right> $ which allow very small spin mixing and hence lead to a lower dephasing of the nuclear spin correlations. A more extensive analysis of various resonant excitation channels at low temperatures for spin selective readout and initialization has been shown by Reiserer et al. \cite{reiserer_robust_2016}

\subsection*{Verification}
Singlet pairing of nuclear spins will affect the way they interact with the central spin. This can then be used as a verification step to know what fraction of the nuclear spin ensemble has been paired into singlets. In Fig. 3 we have shown how singlet pairing can be advantageous in increasing the coherence time of the probe spin which can lead to finer resolution in detecting weakly interacting spins. Though these can be used as signatures for the singlet pairing, there are other direct methods to verify their formation. For this we consider the example of two spins coupled to the central spin by the following Hamiltonian:
\be
H = S^z(g_1 I^x_1 + g_2I^x_2) + \omega(I^z_1 + I^z_2).
\ee
If the free evolution $U(\tau)$ governed by the above Hamiltonian is alternated with $\pi$-pulses on the central spin with the following pulse sequence
\be
U_\frac{\pi}{2} - \left( U(\tau) - U_\pi - U(\tau)\right )^m - U_\frac{\pi}{2}
\ee
where $U_\theta = \exp(-i\theta S^x)$, then a central-spin observable will behave differently if the spins are in the singlet state or in (un)polarized states, i.e. starting from the $S$-spin in state $\ket{0}$, the number of repetitions $m$ required to find it in the state $\ket{1}$ will be
\bea
m &\approx  \frac{\omega}{\sqrt{g_1^2+g_2^2}} &~~~~{\rm unpolarized} \nn \\
m &\approx  \frac{\omega}{|g_1-g_2|} &~~~~ {\rm singlet~ paired}. 
\eea

\subsection*{Calculation of Relaxation}
We introduce dephasing by a Krauss channel, assuming Markovian relaxation such that the state of the central spin is given by
\begin{equation}
\mathcal{E}(\rho(t)) = \frac{1}{2}\left(1-{\rm e}^{-\gamma_d t}\right) \mathbbm{1}\otimes{\rm Tr}_S(\rho(t)) + {\rm e}^{-\gamma_d}\rho(t).
\end{equation}
Here, $ \rho(t) $ is the density matrix of the total system while $ \gamma_d $ is the relaxation rate the central spin is subjected to. $ Tr_S $ defines the partial trace over the central spin.
\subsection*{Simulation of large spin registers}
In order to simulate the evolution of larger ($ N\ge20 $) ensembles of nuclear spins, we make use of the composition of the non-unitary operation $V(\tau) = \frac{1}{2}\left(U^+_I + U^-_I\right)$.  Note first that $H_{\pm}$ in equation \eqref{eq:HPlusMinus} equals the sum of single qubit operations
\begin{equation}
H_\pm = \sum_k \left( \pm \vec{g}_k \cdot \vec{I}_k + \omega I_k^z \right) .
\end{equation}
Successive application of the operation $V$ can therefore be rewritten as 
\begin{equation}
V^n = \dfrac{1}{2^n}(U_1 + U_2)^n = \dfrac{1}{2^n} (\bigotimes_i U_{1,i} +  \bigotimes_j U_{2,j})^n 
\end{equation}
where $U_{1,i}$ is the single qubit unitary acting on the $i$-th qubit. We now rewrite the power $n$ as the sum of all occurring terms and apply it to 
a pure separable input state $ \left| \Psi_0 \right> = \bigotimes_k \left| \Psi_{0,k} \right> $
\begin{eqnarray}
V^n \left| \Psi_0 \right>  =  \dfrac{1}{2^n} \sum_{nPerm_2} \bigotimes_k nPerm_2(U_{1,k},U_{2,k}) \left| \Psi_{0,k} \right> 
\label{eq:MultiPerms}
\end{eqnarray}
Where $ nPerm_2(U_{1,k},U_{2,k}) $ denotes one of the terms obtained when applying the power $n$, more formally the matrix product of a multiset permutation or equally of an $n$-tuple whose entries come from the set $ \{U_{1,k},U_{2,k}\} $ with cardinality 2. 
All of the possible terms are added up and thus the problem is reduced to a sum of Kronecker products of single qubit operations and can be easily parallelized in numerical simulations.
Moreover, the numerical calculations are simplified further by observing that the sum over the permutations in \eqref{eq:MultiPerms} is equivalent to a sum over binary numbers from 0 to $ 2^{n}-1$ when each 0 in the digits of the binary number is replaced by $ U_{1,k} $ and each 1 by $ U_{2,k} $, followed by matrix multiplication.

\section*{Acknowledgements} 

We acknowledge Philipp Neumann, Sebastian Zaiser and Stefan Jesenski for fruitful discussions and comments. We would like to acknowledge the financial support by the Eisele Foundation (J.G.), ERC project SQUTEC, DFG (FOR1693), DFG SFB/TR21, EU DIADEMS, SIQS, Max Planck Society, and the Volkswagenstiftung.
\section*{Author contributions statement}

D.D. conceived the initial concept, J.G. and D.D. conducted the calculations and numerical simulations and wrote the manuscript. J.W. supervised the project. All authors analyzed the results and reviewed the manuscript. 

\section*{Additional information}

\textbf{Competing financial interests}\\ 
The authors declare no competing financial interests. 

\end{multicols}
\end{document}